\documentclass[arXiv]{actapoly}

\begin{document}

\title[Optical transients observed with {\it COLORES}]
{Initial follow-up of optical transients with {\it COLORES} using the {\it BOOTES} network}

\correspondingauthor[M. D. Caballero-Garcia]{M. D. Caballero-Garcia}{1}{cabalma1@fel.cvut.cz}
\author[M. Jelinek]{M. Jelinek}{2}
\author[A. Castro-Tirado]{A. Castro-Tirado}{2}
\author[R. Hudec]{R. Hudec}{1,3}
\author[ ]{R. Cunniffe}{2}
\author[ ]{O. Rabaza}{4}
\author[ ]{L. Sabau-Graziati}{5}

\institution{1}{Czech Technical University in Prague, Technick\'a 2, 166 27  Praha 6 (Prague), Czech Republic}
\institution{2}{Instituto de Astrof\'{\i}sica de Andaluc\'{\i}a (IAA-CSIC), P.O. Box 03004, E-18080, Granada, Spain}
\institution{3}{Astronomical Institute, AS CR, 251 65 Ondrejov, Czech Republic}
\institution{4}{\'Area de Ingenier\'{\i}a El\'ectrica, Dpto. de Ingenier\'{\i}a Civil, Univ. de Granada, Spain}
\institution{5}{Divisi\'on de Ciencias del Espacio (INTA), Torrej\'on de Ardoz, Madrid, Spain}

\begin{abstract}
The Burst Observer and Optical Transient Exploring System (BOOTES) is a
network of telescopes that allows the continuous monitoring of transient astrophysical sources. It was
originally devoted to the study of the optical emission from gamma-ray
bursts (GRBs) that occur in the Universe. In this paper we show the initial
results obtained using the spectrograph COLORES (mounted on BOOTES-2), 
when observing optical transients (OTs) of diverse nature.
\end{abstract}


\maketitle

\section{Introduction}

The majority of the sources visible in the sky have variable emission. Some of them vary in a long time-scale compared to the 
human life. But most of them have shorter variability and, because of that, are continuously observed by telescopes both on 
Earth and space. The study of variability provides one of the most direct clues on the size and physical characteristics
of the emitting object. In the highest energy regime (X-rays, ${\gamma}$-rays) the scales are related to orbits around compact
objects (i.e. the result from the death of massive stars). For longer wave-lengths variability is related to larger regions and/or
bigger stars. The most extreme case is the study of microwaves, that allows the study of large structures in the Universe.
This shows that it is important for astrophysicists to continuously observe our changing Universe, from the smallest to 
the largest scales, for its understanding.

The study of variable sources in the optical regime can be performed from the Earth surface. Furthermore, it is a window for the 
study of many complex physical processes occurring at intermediate space/time scales (see the previous paragraph). In this 
regime we can study (variable) stars
(e.g. erupting variables, novae, cataclysmic variables,... etc.), the most violent physical processes in the Universe (Super-Novae -- hereafter SNe -- and
Gamma-Ray Bursts), the relatively quiet Active Galactic Nuclei (AGN) and Quasars (QSOs) and the outer discs and stellar-companions around compact 
objects (neutron stars and black holes). Nevertheless, the window is still open to discoveries of new types of transients. Recently,
thanks to the advance of devices and detectors, new phenomena are being studied (gravitational lenses, binary mergers,
stellar tidal disruptions by black holes). It is natural to wonder whether there are new (undiscovered) types
of physical processes (sources) giving rise to new kinds of observed emission. 

Observations in the optical are performed by big and medium-sized telescopes on Earth. The former are not suitable for performing the rapid
follow-up needed for the study of optical transients (as we will explain hereafter). These transients events are typically of short duration (from fractions of a 
second to a few days), because the physical processes that originate them are of limited duration/spatial extent. Robotic smaller telescopes are 
very well suited for performing such studies. This is due to several factors: their observing flexibility, their rapid response and
slew times and the fact that they can be located worldwide working remotely (therefore allowing continuous monitoring). Of course, additional 
observations might be triggered after the transient has been detected with large X-ray/Optical Observatories. In this way we can perform
deep studies on the nature of these sources. 

\subsection{The Burst Optical Observer and Transient Exploring System and its Spectrographs}

{\it BOOTES} (acronym of the Burst Observer and Optical Transient Exploring System) is a world-wide network of robotic telescopes. It 
was originally designed from a Spanish-Czech collaboration that started in 1998 (\cite{castro99,castro12}). The telescopes are located 
at Huelva ({\it BOOTES}--1), Malaga ({\it BOOTES}--2), Granada,
Auckland ({\it BOOTES}--3) and Yunnan ({\it BOOTES}--4), located at Spain, New Zealand and China, respectively. There are plans of extending 
this network even further (Mexico, South Africa, Chile,...). These 
telescopes are medium-sized (${\rm D}=30-60$\,cm), autonomous and very versatile. They are very well suited for the continuous study of the fast variability from 
sources of astrophysical origin (Gamma-Ray Bursts -- hereafter GRBs -- and Optical Transients -- hereafter OTs -- ). 

Currently two spectrographs are built and working properly in the network at Malaga and Granada (in the optical and infra-red, respectively). In 
Sec.~\ref{results} we will show preliminary results obtained so far with {\it COLORES} at {\it BOOTES}--2 in the field of OTs of astrophysical origin. 

\subsection{{\it COLORES}}

{\it COLORES} stands for Compact Low Resolution Spectrograph (\cite{rabaza14}). It is a spectrograph designed to be lightweight enough to be carried by the high-speed robotic
telescope 60\,cm ({\it BOOTES}--2). It works in the wavelength range of ($3\,800-11\,500$)\,${\AA}$ and has a spectral resolution of ($15-60$)\,${\AA}$. The 
primary scientific target of the spectrograph is a prompt GRB follow-up, particularly the estimation of redshift.

{\it COLORES} is a multi-mode instrument that can switch from imaging a field (target selection and precise pointing) to spectroscopy by rotating wheel-mounted
grisms, slits and filters. The filters and the grisms (only one is mounted at the moment) are located in standard filter wheels and the optical design is 
comprised of a four-element refractive collimator and an identical four-element refractive camera. As a spectroscope, the instrument can use different 
slits to match the atmospheric seeing, and different grisms in order to select the spectral resolution according to the need of the observation. 

The current detector is a $1\,024{\times}1\,024$ pixels device, with 13 micron pixels. The telescope is a rapid and lightweight design, and a low instrument weight was 
a significant constraint in the design as well as the need to be automatic and autonomous. For further details on description, operation and working with 
{\it COLORES} we refer to M. Jelinek PhD thesis and references therein.

\subsection{Scientific goals} 

The {\it BOOTES} scientific goals are multifold, and are detailed in the following. 

\begin{itemize}
\item{{\bf Observations of GRB optical counterparts}:\\
There have been several GRBs for which the optical counterpart has been detected simultaneously to the gamma-ray event, with magnitudes in the 
range $5^{\rm th}-10^{\rm th}$. These observations provide important results on the central engine of these sources. The fast 
slewing 0.6\,cm {\it BOOTES} telescopes are providing important results in this field (\cite{jelinek10}).
}
\item{{\bf The detection of OTs of cosmic origin}:\\
These events could be unrelated to GRBs and could constitute a new type of astrophysical phenomenon (perhaps associated to QSOs/AGN). If some of them are related
to GRBs, the most recent GRB models predict that here should be a large number of bursting sources in which only transient X-ray/optical emission should be observed,
but no gamma-ray emission. The latter would be confined  a jet-like structure and pointing towards us only in a few cases.
}
\item{{\bf Monitoring a range of astronomical objects}:\\
These are astrophysical objects ranging from galactic sources such as comets, cataclysmic variables, recurrent novae, compact objects in X-ray binaries to extragalactic
sources such as distant SNe and AGN. In the latter case, there are hints that sudden and rapid flares occur.
}
\end{itemize}

\section{Scientific results using {\it BOOTES}-2/COLORES}  \label{results}

In the following we present some important scientific results obtained using {\it BOOTES}-2 and its low-resolution spectrograph {\it COLORES}, since the beginning 
of its operation (Summer of 2012).

\subsection{GRB~130606A}

Since the first light on 1998, more than a hundred of GRBs have been observed with {\it BOOTES}, some of them only ${\approx}30$\,s after the onset of
the gamma-ray event. The majority of the results have been published in circulars and refereed journals \footnote{We refer to:\\
http://bootes.iaa.es  \\
for further information on the {\it BOOTES} network and scientific results.}. Here we focus on the recent event occurred on $6^{\rm th}$ June (2013).

A ${\approx}275$\,s cosmic gamma-ray burst (GRB~130606A) was recorded by {\it Swift} and KONUS-Wind on $6^{\rm th}$ June 2013, 21:04:34
U.T. (T0) (\cite{barthelmy13,golenetskii13}) displaying a bright afterglow (the emission at other wavelengths following the gamma-rays) in X-rays, but no apparent optical
transient emission (\cite{ukwatta13}) in the range of the UVOT telescope aboard {\it Swift}. {\it BOOTES}-2 station automatically responded to the alert 
and an optical counterpart was identified (\cite{jelinek13}),
thanks to the spectral response of the detector up to $1\,{\mu}{\rm m}$, longer than that of {\it Swift}/UVOT (0.17-0.65\,${\mu}{\rm m}$). We refer to \cite{alberto13} for details
on the observations and results.

\subsection{TCP~J17154683-3128303 = NOVA SCORPIUS 2014}

\begin{figure*}
\centering
\includegraphics[bb=14 14 415 399,width=0.8\linewidth]{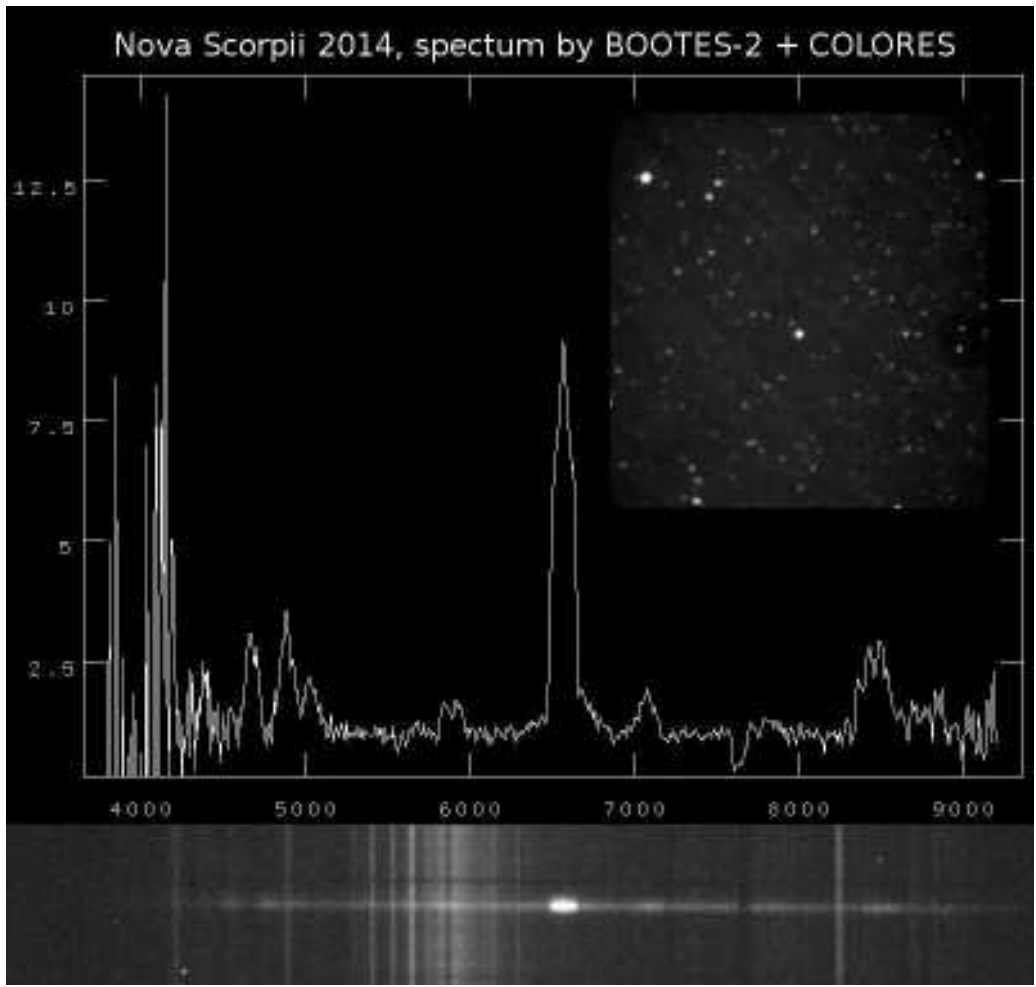} 
\caption{Optical spectrum in the range (3\,800-9\,200)\,${\AA}$ from Nova Scorpii 2014, obtained with {\it BOOTES}-2/COLORES on $30^{\rm th}$ March, 04:37 UT.}
\label{fig1}
\end{figure*}

Following the discovery on $26^{\rm th}$ March (2014) of a $10^{\rm th}$\,mag new source in Scorpius dubbed~TCP J17154683-3128303 by \cite{nishiyama14} 
also detected by {\it Swift} on Mar 27 (\cite{kuulkers14}), \cite{jelinek14} report that an 
optical spectrum was obtained 
with the COLORES spectrograph at the 0.6\,m robotic telescope at the {\it BOOTES}-2 astronomical station (see Figure~\ref{fig1}).
The spectrum, covering the range (3\,800-9\,200)\,${\AA}$ has been taken on $30^{\rm th}$ March, 04:37 UT and shows broad emission lines of Balmer series, He~I 501.6, 587.8, 706.5, 
and probably of O~I 844.6, suggesting a nova in early phase (Nova Scorpius 2014) thus confirming the earlier suggestion by 
\cite{noguchi14}.

\subsection{Optical spectra from red-dwarf flaring stars}

The {\it Swift} team reported on the detection of a superflare from one of the stars
in the close visual dM4e+dM4e flare star binary system DG~CVn.
The Burst Alert Telescope (BAT) triggered on DG~CVn the
$23^{\rm rd}$ April (2014) at 21:07:08 UT = T0 (\cite{delia14}). {\it BOOTES}-2/COLORES has been observing DG~CVn since 
the beginning of the superflare and new interesting optical spectral variability following the X-ray superflare evolution 
has been observed (we refer to \cite{caballero14} for further details).

\begin{figure*}
\centering
\includegraphics[bb=22 16 727 595,width=0.8\linewidth]{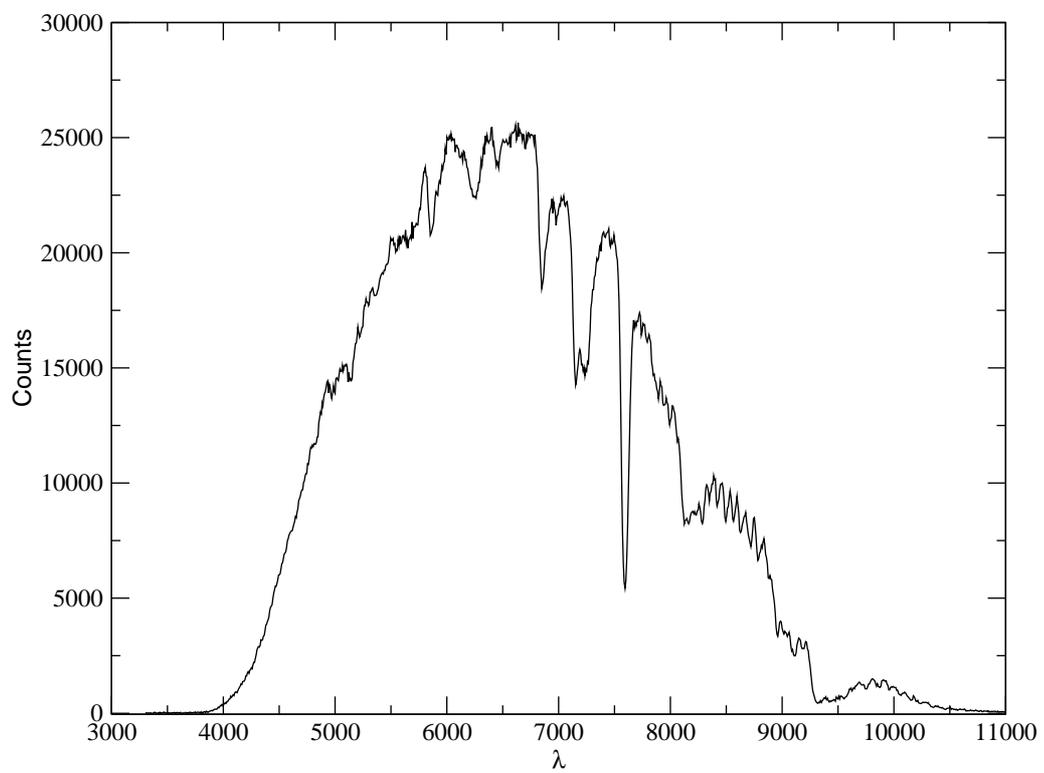} 
\caption{Optical spectrum in the range 3\,800-11\,500\,${\AA}$ from UX~Ari, obtained with {\it BOOTES}-2/COLORES on $19^{\rm th}$ July, 01:32 UT.}
\label{fig2}
\end{figure*}

Additionally, following the detection and subsequent monitoring of the new outburst from the RS CVn UX~Ari by SWIFT and MAXI (\cite{kawagoe14,krimm14}), 
the 0.6\,m robotic telescope at the {\it BOOTES}-2 astronomical station, obtained optical (4\,000-9\,000)\,${\AA}$ spectra starting at 
$19^{\rm th}$ July (2014), 01:32:24.382 UT and ending at 04:25:55.652 UT. \cite{caballero14b} report that the optical spectra contain broad 
molecular TiO, CaI, MgI, NaI lines plus a red continuum (see Figure~\ref{fig2}). These spectra lack of any significant Balmer 
lines in emission. These spectral features are indicative of a late-type star spectrum (as previously reported). Nevertheless, 
there are no indications of important chromospheric activity, that might have been disappeared by the time of our observations.

\section{Discussion and conclusions}

The era of OTs is about to start. Since the beginning of the modern times the big telescopes have been the only
resource for astronomers to study astrophysical sources. In spite of constituting the best tool for deep studies of individual
targets, they are not properly suited for the discovery of optical transient sources. Their big size limits the speed they can 
cover the entire sky and the time for overheads might be longer than that for medium-sized telescopes. Many factors make them difficult to
fully automatize and indeed currently none is completely robotic and autonomous. Medium-sized telescopes (i.e. ${\rm D}{\leq}1$\,m) can be 
much quicker moving from target to target and time overheads are usually very small. Therefore, robotic medium-sized telescopes
are currently the best ones for the follow-up and studies of the long-term variability of the astrophysical transient sources.

{\it BOOTES}-2 constitutes one step forward with respect to any robotic telescope on Earth that has existed so far. This is because
it is the {\it first robotic telescope with a spectrograph} mounted on it which has demonstrated to work properly. This 
does not allow only to perform early follow-up and to measure
the redshift of GRBs, with cosmic origin, but also to perform early follow-up (both photometric and spectroscopic) of transient sources, 
often located much
more nearby. These OTs might have been reported at other wave-lengths (typically at X- and ${\gamma}$-rays), which would
create an alert to the scientific community (often through an {\it Astronomer Telegram}). In such a case the observer (remotely) sends the telecommands to start
the observation with {\it BOOTES}-2, that re-observes the target every time it is visible during the following nights.

Apart from the intensive campaign of follow-up of GRBs performed by the {\it BOOTES} network (${\approx}100$ GRBs have been observed so far), {\it BOOTES}-2
and its spectrograph (COLORES) are providing excellent results in the field of OTs, too. In this paper we mention a few of them, obtained during the last
1.5\,year (since the spectrograph was mounted on the telescope). But this is only the beginning and we look forward to follow
many OTs for understanding better their physical properties and may be also to discover and follow-up new kinds of OTs. This
will prepare us for the advent of the {\it Large Synoptic Survey Telescope} (LSST), the biggest telescope ever built on Earth for the study of the 
entire sky, that will allow the discovery of many kinds of new astrophysical sources (if they exist) and follow-up of OTs, planned to start operations
on 2023.

\begin{acknowledgements}
MCG is supported by the European social fund within the framework 
of realizing the project "Support of inter-sectoral mobility and quality enhancement 
of research teams at Czech Technical University in Prague", CZ.1.07/2.3.00/30.0034. MJ 
and AJCT thank the support of the Spanish Ministry Projects AYA2009-14000-C03-01 and 
AYA2012-39727-C03-01. RH acknowledges GA CR grant 13-33324S.
\end{acknowledgements}

\bibliographystyle{actapoly}
\bibliography{paperibws}

\end{document}